\documentclass[%
 reprint,
superscriptaddress,
 amsmath,amssymb,
 aps,
prb,
floatfix
]{revtex4-2}

\usepackage{graphicx}% Include figure files
\usepackage{dcolumn}% Align table columns on decimal point
\usepackage{bm}% bold math
\usepackage{units}
\usepackage{hyperref}% add hypertext capabilities
\usepackage{xcolor}
\usepackage{xspace}
\usepackage{mathbbol}

\newcommand{\ahfx}{\ensuremath{\alpha_\text{HFX}}\xspace}

\begin{document}

\title{Uncertainty Quantification in Multiscale Models of Charge Transport in Organic Semiconductors: Influence of the Exhange-Correlation Functional}

\author{Zhongquan Chen}
\author{Pim van der Hoorn}
\author{Bj\"orn Baumeier}
\affiliation{Department of Mathematics and Computer Science \& Institute for Complex Molecular Systems, Eindhoven University of Technology}

\date{\today}

\begin{abstract}
This study investigates the impact of exchange-correlation functional choices on the predictive accuracy of multiscale models for charge transport in organic semiconductors (OSCs). A hybrid functional approach is applied to analyze uncertainties in key parameters influencing charge mobility, focusing on the Hartree--Fock exchange fraction. Using 2-methyl-9,10-bis(naphthalen-2-yl)anthracene (MADN) as a test system, molecular dynamics and density functional theory are combined to compute reorganization energies, site energies, and electronic coupling elements. Monte Carlo sampling quantifies the uncertainty propagation, revealing that site energy variations dominate transport property uncertainties, while coupling elements contribute minimally. The findings underscore the need for accurate parameter determination and functional selection, with implications for enhancing the reliability of first-principles-based multiscale modeling frameworks in OSC design.  
\end{abstract}

\maketitle

\section{Introduction}
Organic semiconductors (OSCs) are materials that consist of organic molecules, often in disordered structures formed during spin-coating or deposition processing. Besides their semiconducting properties, they are mechanically flexible and come with the potential to control charge transport properties~\cite{hamers_flexible_2001,liu_high_2015,chow_organic_2020}. This flexible functionality is achieved by tuning molecular properties to achieve an ideal operation at the device level~\cite{bronstein_role_2020, bredas_organic_2002}. Computational approaches that aim at predicting the charge transport properties of disordered molecular materials resolving the interplay between single-molecule properties (such as their electronic structure or response properties) and mesoscale material morphology can play an important role in supporting or even guiding experimental optimization processes~\cite{bronstein_role_2020, sokolov_computational_2011, grynova_read_2018}.

\begin{figure*}[tbp]
  \centering
  \includegraphics[width=\linewidth]{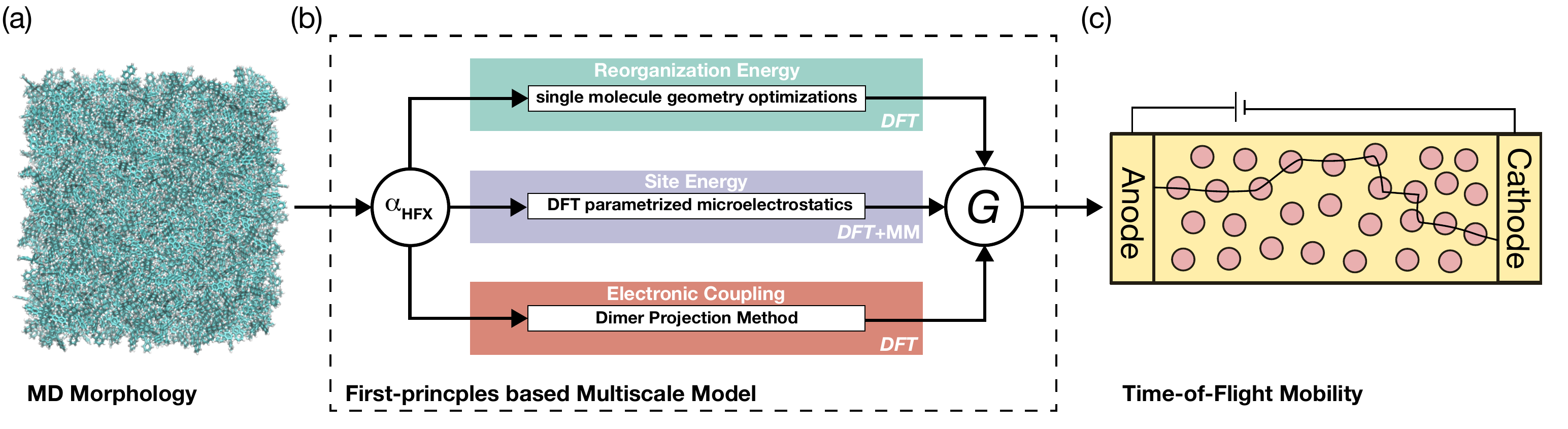}
  \caption{Schematic representation of the multiscale framework for simulation of charge transport: (a) a representative large-scale morphology is obtained from Molecular Dynamics. (b) Information about the molecular building blocks and the morphology are used to determine the molecular transport graph $\mathbf{G}$ for a given choice of the \ahfx via calculation of reorganization energy, site energies, and electronic couplings. (c) The time-of-flight charge mobility is simulated based on the graph.}
  \label{fig:MSM}
\end{figure*}

First-principles based multiscale models are such types of computational approaches. Due to the infeasibility of explicitly simulating the coupled non-adiabatic electron-nuclear dynamics for the time- and length scales of realistic materials, these models typically exploit the localization of electronic states in disordered molecular materials and consider a rate-based description of hopping-type transport. A popular choice for the electron transfer rate $\omega_{ij}$ between two localization sites $i$ and $j$ is Marcus theory~\cite{marcus_theory_1956, marcus_electron_1993}, in which
\begin{equation}
    \omega_{ij} = \frac{2\pi}{\hbar} \frac{|J_{ij}|^2}{\sqrt{4\pi \lambda_{ij} k_\text{B}T}} \exp\left(-\frac{(\Delta E_{ij} - \lambda)^2}{4\lambda k_\text{B}T}\right) ,
    \label{equ:Marcus}
\end{equation}
where $\hbar$ is the reduced Planck constant,  $k_\text{B}$ the Boltzmann constant, and $T$ the temperature. As depicted in Fig.~\ref{fig:MSM}, given a larger scale morphology and with the definition of localization sites (Fig.~\ref{fig:MSM}(a)), multiscale models employ first-principles based methods (Fig.~\ref{fig:MSM}(b)) to explicitly calculate the remaining physical, material-specific (or rather transfer-pair-specific) quantities in Eq.~\ref{equ:Marcus}: the reorganization energy $\lambda$ (here a single value as we assume a single-component material), the electronic coupling $J_{ij}$, and the site energy difference $\Delta E_{ij} = E_i - E_j$. With all that information, charge transport is modeled as a continuous time random walk (CTRW) process on a graph $\mathbf{G}$, constructed from the localization sites and the calculated rates between them (Fig.~\ref{fig:MSM}(c)). Panel (b) of Fig.~\ref{fig:MSM} mentions some specific methods for the calculation of the quantities entering Eq.~\ref{equ:Marcus}, such as mixed quantum-classical methods for obtaining the site energies with the help of microelectrostatic methods~\cite{poelking_impact_2015, poelking_long-range_2016} or the dimer projection method for determining coupling elements~\cite{baumeier_density_2010}, as they are the ones adopted in this work. What is essential about these and alternative ones is that they typically rely on density-functional theory (DFT) calculations, either directly or as a mean to parametrize classical models. The dependence of DFT calculations on the choice of an exchange-correlation functional raises the question of how sensitive the simulated charge transport is to this choice and how certain predictions of material properties are. 

Uncertainty quantification (UQ) is concept from computational science which allows for an estimation of confidence intervals for a quantity of interest (QoI) and an analysis of its sensitivity in models of a physical system that contain uncertain, maybe empirical, or noisy, parameters~\cite{sarkar_uncertainty_2017,oconnor_quantifying_2024, chernatynskiy_uncertainty_2013, suleimenova_tutorial_2021,coveney_reliability_2021, coveney_when_2021}. Many common UQ studies focus on models with (partial) differential equations, e.g., drift-diffusion equations in which the diffusivity as parameter, and assume a certain distribution for the values of the parameter(s). For the multiscale model of charge transport, it is not straightforward to cast the large variety of available exchange-correlation functionals into the role of a model parameter with some distribution. To keep the problem tractable, we focus here instead specifically on the exchange part in hybrid functionals~\cite{perdew_rationale_1996,marques_densitybased_2011}, in which a DFT model for the exchange is mixed using a weighting factor $\alpha_\text{HFX}$ with a Hartree--Fock type exchange, i.e.,
\begin{equation}
  E_\text{x} = \alpha_\text{HFX} E_\text{x}^\text{HF} + (1-\alpha_\text{HFX})E_\text{x}^\text{DFT}.
  \label{equ:hybrid}
\end{equation}
More specifically, we take as the basis the PBE0 functional~\cite{adamo_toward_1999} and scrutinize (i) how the predictions of the multiscale model of charge transport are affected by variation of $\alpha_\text{HFX}$ as a proxy for uncertainties in the choice of DFT functionals, (ii) what the level of confidence is in quantitative predictions, and (iii) what are the most sensitive quantities in the model. In this sense, Eq.~\ref{equ:hybrid} is deceptively simple. For each value of $\alpha_\text{HFX}$, the graph $G$ is constructed using the respective value of the reorganization energy, the $N_\text{mol}$ site energies, and the $N_\text{pair}$ coupling elements, and the dimensionality of the problem from the perspective of UQ is $N_\text{UQ}=1+N_\text{mol}+N_\text{pair}$, which can easily be on the order of $10^{4}-10^{5}$. We consider the simulated time-of-flight (ToF) and the associated mobility the QoI in the following (see Fig.~\ref{fig:MSM}(c)) which then are subject to uncertainties in these $N_\text{UQ}$ parameters, {\em stemming} from the variation in $\alpha_\text{HFX}$. As a prototypical system, we will study hole transport in an amorphous morphology of 2-methyl-9, 10-bis(naphthalen-2-yl)anthracene (MADN), a wide-gap semiconductor that is used extensively as an ambipolar host material in organic light-emitting diodes~\cite{ko_accurate_2019, chang_great_2017}. 

This paper is organized as follows. Section~\ref{sec:model} outlines the theoretical and computational details of the multiscale model, including the morphology simulation with classical molecular dynamics, the determination of all quantities in the transition rates, and the calculation of the charge transport properties from the constructed graph model. In Section~\ref{sec:MSMresults}, we discuss the explicit results from the model using different values of $\alpha_\text{HFX}$, before we show the results of uncertainty quantification and sensitivity analysis via Monte Carlo sampling in Section~\ref{sec:UQ}. A brief summary concludes the paper. 

\section{Multiscale Model}
\label{sec:model}
The multiscale model maps a large scale molecular morphology with atomistic detail into a graph $\mathbf{G}(\mathbf{V}, \mathbf{W})$, where the set of nodes $\mathbf{V}$ is determined from the center-of-masses of the individual molecules and $\mathbf{W}$ is the adjacency matrix formed by the Marcus rates $\omega_{ij}$. Two nodes $i,j$ are connected if the corresponding molecules have their closest-contact distance smaller than $r_\text{cutoff}=\unit[0.5]{nm}$. 

\subsection{Molecular Dynamics}
Classical molecular dynamics (MD) is used to create an amorphous morphology of MADN. An empirical force-field for these simulations has been obtained via the Automated Topology Builder~\cite{stroet_automated_2018}, and an initial structure containing 1000 molecules in a cubic cell is created. Periodic boundary conditions are applied throughout in all three spatial directions. 
After energy minimization, the system is simulated for \unit[1]{ns} in the $NpT$ ensemble, keeping a constant temperature of \unit[300]{K} and constant pressure of \unit[1]{bar} using the velocity-rescale thermostat~\cite{bussi_canonical_2007} with the coupling time constant \unit[0.1]{ps} and the Parrinello-Rahman barostat~\cite{parrinello_polymorphic_1981} with a time constant for pressure coupling \unit[2]{ps}. The equation of motion for updating the atomic coordinates is implemented by leap-frog algorithm~\cite{van_gunsteren_leap} with a time step of \unit[1]{fs}. Following this, the temperature is increased to \unit[800]{K}, well above the glass transition temperature, during a period of \unit[0.5]{ns}. This temperature is maintained for \unit[1]{ns} before cooling back down to \unit[300]{K} during a period of \unit[0.5]{ns}. Such a heating-cooling cycle is repeated three times. After this simulated annealing, a production run is conducted for \unit[2]{ns} using the $NpT$ ensemble. The final configuration of MADN is chosen for the further steps in the multiscale model, whose configuration is a cubic box with a length of \unit[9.0]{nm} and a density of $\unit[1.08]{g/cm^3}$. All calculations have been performed with the GROMACS software package~\cite{berendsen_gromacs_1995}.

\subsection{Electronic Structure Calculations} 
\label{sec:es}
Molecular orbitals $\phi_l (\mathbf{r})$ with energies $\epsilon_l$ of the individual molecules in the morphology are obtained within DFT as the solutions to the  Kohn--Sham equations~\cite{kohn_self_1965}
\begin{eqnarray}
    && \left(-\frac{1}{2}\nabla^2_{\mathbf{r}} + v_\text{ext}(\mathbf{r}) + v_\text{H}[\rho](\mathbf{r}) + v_\text{XC}[\rho](\mathbf{r})\right) \phi_l(\mathbf{r}) \nonumber \\
    && = H^\text{KS} \phi_l(\mathbf{r}) = \epsilon_l \phi_l (\mathbf{r}) ,
    \label{eq:KS2}
\end{eqnarray}
where $v_\text{ext}$ is an external potential (typically from the nuclei), $v_\text{H}[\rho]$ the electrostatic Hartree potential of a classical charge density $\rho(\mathbf{r})$, and $v_\text{XC}[\rho]$ the exchange-correlation potential containing explicit quantum-mechanical electron-electron interactions. The charge density is determined from the single-particle wave functions as $\rho(\mathbf{r})=\sum\limits_{l=1}^{N_\text{el}} \left\vert\phi_l(\mathbf{r})\right\vert^2$. As the Hartree and exchange-correlation potential depend on this density, solutions to Eq.~\ref{eq:KS2} have to be found self-consistently. This corresponds to finding the ground-state density $\rho_0$ that minimizes the total energy of the system expressed as
\begin{equation}
    U=U[\rho] = T_s[\rho] + \int v_\text{ext}(\mathbf{r}) \rho(\mathbf{r}) d \vec{r} + E_\text{H}[\rho] + E_\text{XC}[\rho]
    \label{eq:KS_model}
\end{equation}
where $T_s[\rho]$ is the kinetic energy, $E_\text{H}[\rho]$ and $E_\text{XC}[\rho]$ the Hartree and exchange-correlation energies, respectively. 

The practical calculations in this work have been performed with the ORCA software~\cite{Neese2012a} using the def2-tzvp~\cite{weigend_accurate_2006} basis set to represent the Kohn--Sham wave functions. As mentioned in the Introduction, the correlation part of the exchange-correlation functional is taken from the PBE0 functional, while the weighting factor of the Hartree--Fock type exchange in the exchange part is varied. In common practice, $\alpha$ is small and below 0.25. For our uncertain quantification study, $\alpha=0,0.05,0.10,0.15,0.20,0.25$ are chosen for the multiscale model.

\subsection{Reorganiztion Energy}
 The reorganization energy $\lambda$ accounts for the energy change caused by the geometry variation during the charge transport, and is linked to four points on the potential energy surfaces of neutral (n) and charged (c) molecules at neutral (N) or charged (C) equilibrium geometries via:
\begin{equation}
    \lambda_{ij} = U_i^\text{nC} - U_i^\text{nN} + U_j^\text{cN} - U_j^\text{cC},
    \label{eq:lambda}
\end{equation}
where $U^\text{xX}$ is the total DFT energy of $\text{x}=\text{n},\text{c}$ state in the $\text{X}=\text{N},\text{C}$ geometry. While in principle transfer pair specific, we use a single value of all molecular pairs.

\subsection{Site Energy}
The site energy $E_i = E_i^\text{c} - E_i^\text{n}$ is the difference between the total energies of the system in which molecule $i$ is carrying a charge or not, corresponding to the ionization potential in case of hole transport and the negative of the electron affinity in case of electron transport. The individual total energies in turn consist of different contributions associated with different physical mechanism, i.e.,
\begin{equation}
E_i^x = U_i^\text{xX} + E_i^{\text{x},\text{el}} + E_i^{\text{x},\text{polar}},
\label{eq:Es}
\end{equation}
where $U_i^\text{xX}$ is the internal energy contributions and both $E_i^{\text{x},\text{el}}$ and $E_i^{\text{x},\text{polar}}$ are contributions arising from purely static and polarizable intermolecular interactions, respectively. As those interactions are typically long-ranged, the intermolecular contributions to the site energy can typically not be calculated with a fully quantum-mechanical method and classical models are adopted, instead, which we refer to as a microelectrostatic model using moment representations parametrized based on single molecule DFT reference data. Specifically, in the multiscale model here, we employ a point charge respresentation~\cite{jcc540110311} for the electrostatic potential of charged and neutral molecules, so that the electrostatic energy contribution is
\begin{equation}
    E_i^{\text{x},\text{el}} = \frac{1}{4 \pi \epsilon_0} \sum\limits_{a_i} \sum\limits_{b_k,k \neq i} \frac{q^\text{x}_{a_i}q^n_{b_k}}{ |\mathbf{R}_{a_i} - \mathbf{R}_{b_k}|} 
\end{equation}
where $\epsilon_0$ is the vacuum permittivity, $a_i, b_k$ denotes the atoms in molecule $i,k$, $q^{\text{x}}_{a_i}$ are the partial charge of atom $a$ when molecule $i$ is in state $\text{x}$. To account for effects of polarization, and to evaluate $ E_i^{\text{x},\text{polar}}$, we use the model of distributed atomic dipole polarizabilities (Thole model)~\cite{thole_molecular_1981}, in which the parameters are also determined such that the classical volume of the molecular polarizability tensor matches the DFT reference.
Intermoelcular effects are considered in a region of \unit[4.0]{nm} around each individual molecule. Practical calculations of the site energies are performed using the VOTCA software~\cite{Baumeier2011,doi:10.1021/acs.jctc.8b00617,10.1063/1.5144277,Baumeier2024}.

\subsection{Electronic Coupling Elements}
The coupling element $J_{ij}$ between molecule $i$ and $j$ describe the coupling strength between two localized states, here approximated by monomer single-particle wavefunctions $\vert \phi_i\rangle$ and $\vert\phi_j\rangle$, respectively. For hole transport, the relevant orbitals are highest-occupied molecular orbital (HOMO). Using the Dimer-Projection Method~\cite{baumeier_density_2010} the coupling element is determined as:
\begin{equation}
    J_{ij} = \frac{ J^0_{ij}- \frac{1}{2}(e_i+e_j) S_{ij} }{ 1- S_{ij}^2 }
    \label{equ:JAB}
\end{equation}
where $J^0_{ij} = \langle \phi_i | \hat{H}^\text{KS}_\text{D} | \phi_j \rangle $, $e_i = \langle \phi_i | \hat{H}^\text{KS}_\text{D} | \phi_i \rangle $, $e_j = \langle \phi_j | \hat{H}^\text{KS}_\text{D} | \phi_j \rangle $, and $S_{ij}=\langle \phi_i | \phi_j \rangle $ with bra-ket notation. The Hamiltonian of the dimer,  $H^\text{KS}_\text{D}$ (see Eq.~\ref{eq:KS2}), is diagonal in its eigenbasis $\left\{\vert \phi^\text{D}_k\rangle\right\}$ with eigenvalues $\left\{ \epsilon^\text{D}_k\right\}$, so $H^\text{KS}_\text{D} = \text{diag}(\epsilon^\text{D})$. With the projections of the monomer functions on the dimer eigenbasis, i.e., $p_{ik} = \langle \phi_i | \phi^\text{D}_k \rangle$ and  $p_{jk} = \langle \phi_j | \phi^\text{D}_k \rangle$, $J^0_{ij}$ can be calculated as $J^0_{ij} = \mathbf{p}_i^\text{T} \text{diag}(\epsilon^\text{D}) \mathbf{p}_j$. Similarly, $e_{i(j)} = \mathbf{p}_{i(j)}^\text{T} \mathbf{p}_{i(j)}$ and $S_{ij} =  \mathbf{p}_i^\text{T} \mathbf{p}_j$. All  of these operations are performed in the basis set representation of the Kohn--Sham wave functions (see Sec.~\ref{sec:es}) as implemented in VOTCA.

\subsection{Time-of-flight Calculation}
After constructing the  graph $\mathbf{G}$ from the multiscale model, charge dynamics are modeled as a continuous-time random walk on this graph. In the time-of-flight (ToF) model, some vertices serve as source nodes, representing the electrode where charge carriers are injected, and some as sink nodes, where charge carriers are detected and the ToF is recorded. In CTRW the ToF is calculated as the expected hitting time of a continuous time Markov chain. For a system with $N$ molecules and one charge carrier, the transition rates between the molecules define the adjacency matrix $\mathbf{W}$:
\begin{equation}
  \label{eq:transition_rates}
	\mathbf{W}_{ij} =
	\begin{cases}
	     0			&  i \text{ is not connected to } j,\\
         \omega_{ij}   &  i \text{ is connected to } j,
	\end{cases}
\end{equation}

Then the transition probability from state $i$ to $j$ is $p_{ij} = \mathbf{W}_{ij}/D_i$ where $D_i := \sum_{j} \mathbf{W}_{ij}$.
And the expected time from node $i$ to reach the sink $\tau_i$ is calculated via: 
\begin{equation}
  \label{eq:hitting_time}
	\tau_i = \begin{cases}
		\frac{1}{D_i} + \sum_{j \ne i} p_{ij} \tau_{j} &\text{if $i$ is not a sink node},\\
		0 &\text{else.} 
	\end{cases}
\end{equation} 

To account for all possible starting nodes of the carrier, all source nodes must be considered. The random walk process can be modeled as a parallel electric network of capacitors~\cite{doyle_random_1984}. Accordingly, the ToF is evaluated using the harmonic mean:
\begin{equation} 
  \tau = N_\text{source} \left[\sum_{i \in \text{Source}} (\tau_i)^{-1}\right]^{-1},
  \label{eq:ToF}
\end{equation}
where $N_\text{source}$ is the number of source nodes.

To determine the ToF in the simulated MADN system along the positive $x$-direction, we remove the PBC in this direction and define the nodes with coordinates $\unit[0]{nm} < x < \unit[0.5]{nm}$ as source nodes, and those with coordinates $\unit[8.5]{nm} < x < \unit[9]{nm}$ as sink nodes. These definitions are changed accordingly for simulating transport in negative $x$-direction, and $y$- or $z$-directions.

\section{Explicit Results from the Multiscale Model}
\label{sec:MSMresults}
In this section, we present and analyze the explicit results of the multiscale model of charge transport in amorphous MADN as obtained for different values of $\alpha_\text{HFX}$.

\subsection{Molecular Parameters}

\begin{figure}[tbp]
    \centering
    \includegraphics[width=\linewidth]{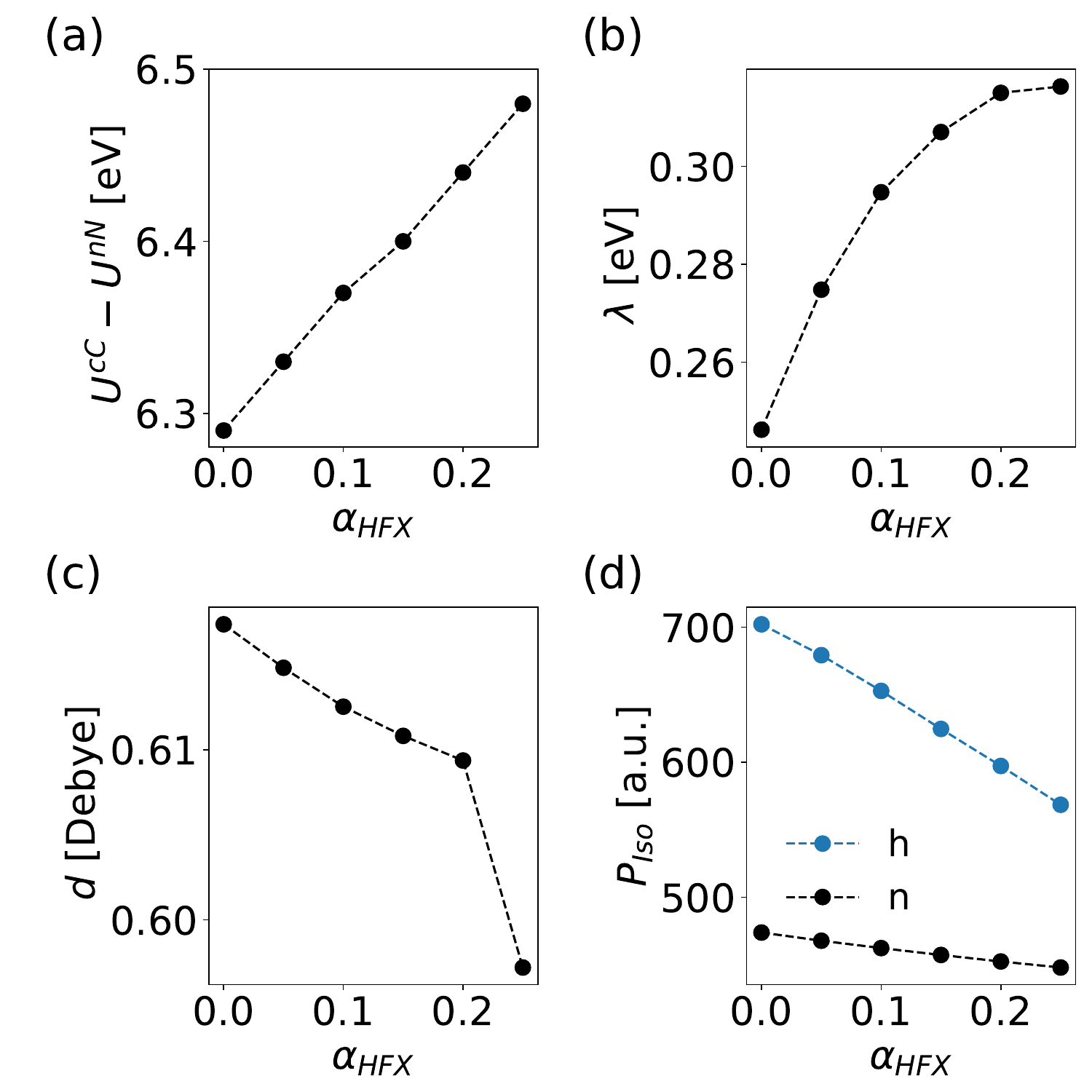}
    \caption{Dependence of molecular parameters as used directly in the multiscale model or in its paramaterization phase on the amount of Hartree--Fock type exchange in the PBE0-based hybrid functional $\alpha_\text{HFX}$. (a) The adiabatic ionization potential, (b) reorganization energy, (c) dipole moment of the neutral molecule, and (d) isotropic molecular polarizability in neutral and charged states, respectively.}
    \label{fig:autogen_MADN}
\end{figure}

We begin with a brief discussion of the molecular parameters as they are used in different ways in the multiscale model. Figure~\ref{fig:autogen_MADN} shows the adiabatic ionization potential, reorganization energy, neutral state dipole moment, and isotropic polarizability in neutral and cationic (hole) states, respectively. The adiabatic ionization potential in Fig.~\ref{fig:autogen_MADN}(a) in principle contributes to the site energy but as it is determined per molecule-type in the system, it has no effect on the site-energy difference $\Delta E_{ij}$ in Eq.~\ref{equ:Marcus}. It is nevertheless interesting to see that it increases almost linearly over the shown range of \ahfx. In contrast, the reorganization energy as shown in Fig.~\ref{fig:autogen_MADN}(b) appears to saturate for \ahfx after an initially close to linear increase. In total, $\lambda$ is found to be in an interval between \unit[0.25]{eV} and \unit[0.33]{eV}. Panels (c) and (d) of Fig.~\ref{fig:autogen_MADN} show the dipole moment of the neutral MADN molecule and the isotropic polarizability of the neutral and cationic (hole) states, respectively, both as electrostatic properties that enter indirectly the parameterization of the microelectrostatic model. As is visible, the dipole moment is rather independent on \ahfx (note that the jump of the last shown data point appears more pronounced because of the very small scale on the $y$-axis). The isotropic polarizabilities in panel (d) exhibit a linear decrease with increasing value of \ahfx which can be attributed to an increasingly attractive effective potential from stronger the Hartree--Fock-like exchange term and consequently more strongly bound electrons as one can also see from the increasing ionization potential in panel (a).  

\subsection{Site energies}

\begin{figure*}[tbp]
  \centering
  \includegraphics[width=\linewidth]{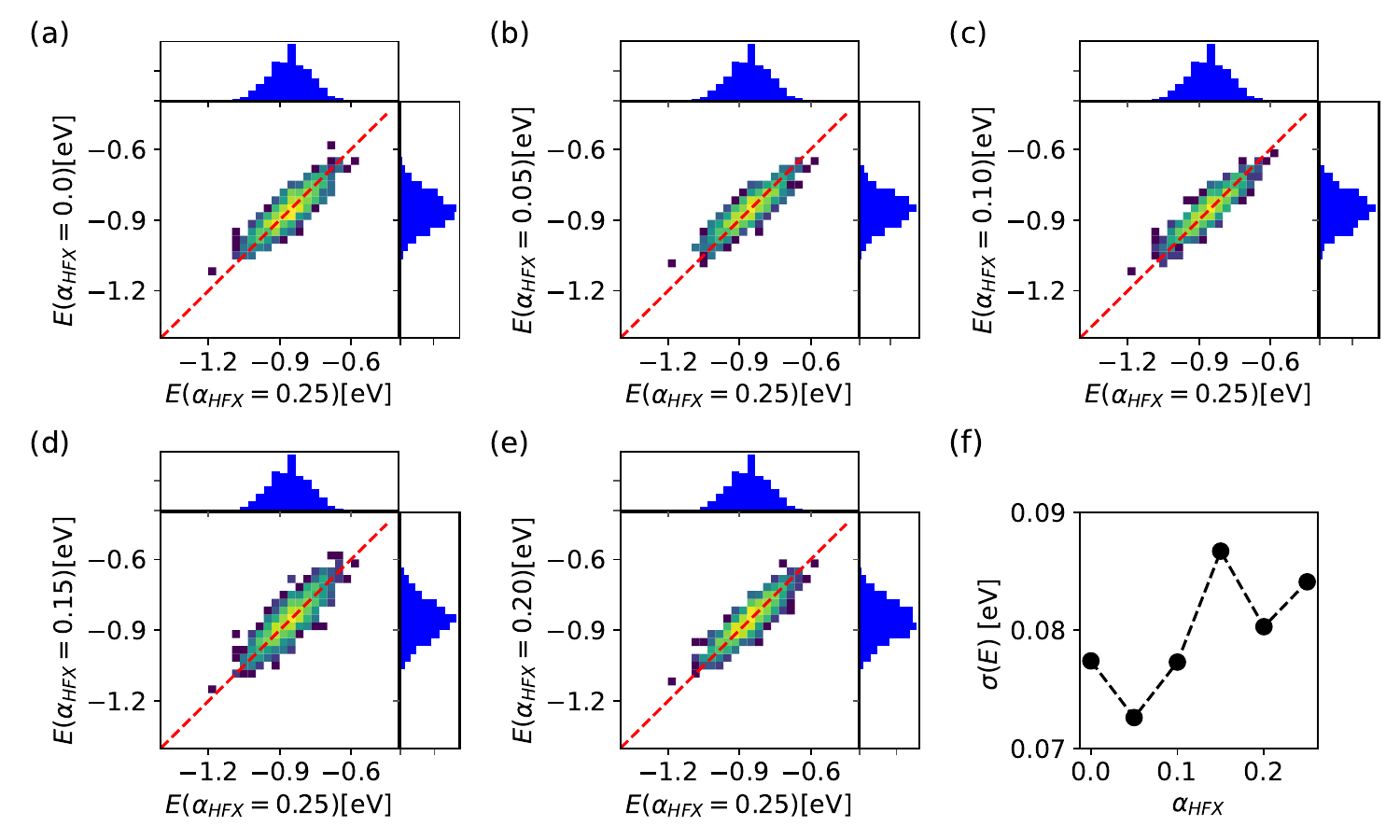}
  \caption{Panels (a)-(e): Scatter plot of site energies of the MADN system for different values of \ahfx vs the PBE0 ($\ahfx=0.25$) reference. (f) Site energy disorder $\sigma$ in \unit[]{eV} for all values of \ahfx.}
  \label{fig:E_qmmm_MADN}
\end{figure*}

\begin{table}[tbp]
  \caption{Characteristics of site energies $E$ (all in \unit[]{eV}) obtained for the different values of \ahfx : minimum ($E_\text{min}$) and maximum ($E_\text{max}$), the overall standard deviation and the contributions for static and polar calculations, as well as the largest absolute variation $D_{\ahfx}$ with respect to the PBE0 data ($\ahfx=0.25$).}
  \begin{ruledtabular}
    \begin{tabular}{c c c c c c c}
    $\ahfx$ & $E_\text{min}$ & $E_\text{max}$ & $\sigma$ & $\sigma_\text{el}$ & $\sigma_\text{polar}$ &  $D_{\ahfx}$ \\
    \hline
    0    & -1.11 & -0.59 & 0.08 & 0.10 & 0.04 & 0.13 \\ 
    0.05 & -1.07 & -0.63 & 0.07 & 0.09 & 0.04 & 0.12 \\ 
    0.10 & -1.13 & -0.61 & 0.08 & 0.10 & 0.04 & 0.13 \\ 
    0.15 & -1.16 & -0.59 & 0.09 & 0.11 & 0.05 & 0.15 \\ 
    0.20 & -1.12 & -0.63 & 0.08 & 0.10 & 0.05 & 0.12 \\ 
    0.25 & -1.17 & -0.60 & 0.08 & 0.11 & 0.04 & ---  \\ 
    \end{tabular}
  \end{ruledtabular}
  \label{tab:maxEi}
\end{table}

Figure~\ref{fig:E_qmmm_MADN}(a)-(e) presents scatter plots comparing the distributions of the calculated site energies of MADN molecules for different \ahfx values, using $\ahfx = \unit[0.25]{}$ (PBE0 functional) as a reference in all cases. Visually, the distributions look very similar as most data points cluster near the diagonal line. In Table~\ref{tab:maxEi}, we report the minimum ($E_\text{min}$) and maximum ($E_\text{max}$) values of the energies and the standard deviation $\sigma$ of the individual distributions, which are also shown in Fig.~\ref{fig:E_qmmm_MADN}(f). All noted variations in these quantities appear small. The data in Figures~\ref{fig:Estat_qmmm_MADN} and \ref{fig:Edip_qmmm_MADN} in the Appendix show that the observations on the total site energies also hold for the individual contributions of electrostatic and polarization energies. This is also reflected by the small variations in the respective standard deviations $\sigma_\text{el}$ and $\sigma_\text{polar}$ listed in Tab.~\ref{tab:maxEi}.

Interestingly, we find the biggest variation of \unit[0.1]{eV} in the values for $E_\text{min}$. This minimal energy (or rather all the energies in the low-energy region of the distributions) can have a very significant influence on the charge transport properties, depending on how much the site energy differences are influenced in the charge transport network. To further characterize the effect of different \ahfx values, we consider the largest absolute variation in the site energies depending on \ahfx by calculating the Chebyshev distance $D_{\ahfx} = \max\limits_{i}(|E_i(\ahfx=0.25) - E_i(\ahfx)|)$. From the results listed in Tab.~\ref{tab:maxEi}, one can see that on individual level the site energies can vary substantially, with more than \unit[0.1]{eV}. This variation is larger than the recorded values of $\sigma$ and with all \ahfx yielding the same mean site energy of \unit[-0.86]{eV} this could indicate non-neglibible effects on charge transport. 

\subsection{Electronic couplings}

\begin{figure*}[tbp]
  \centering
  \includegraphics[width=\linewidth]{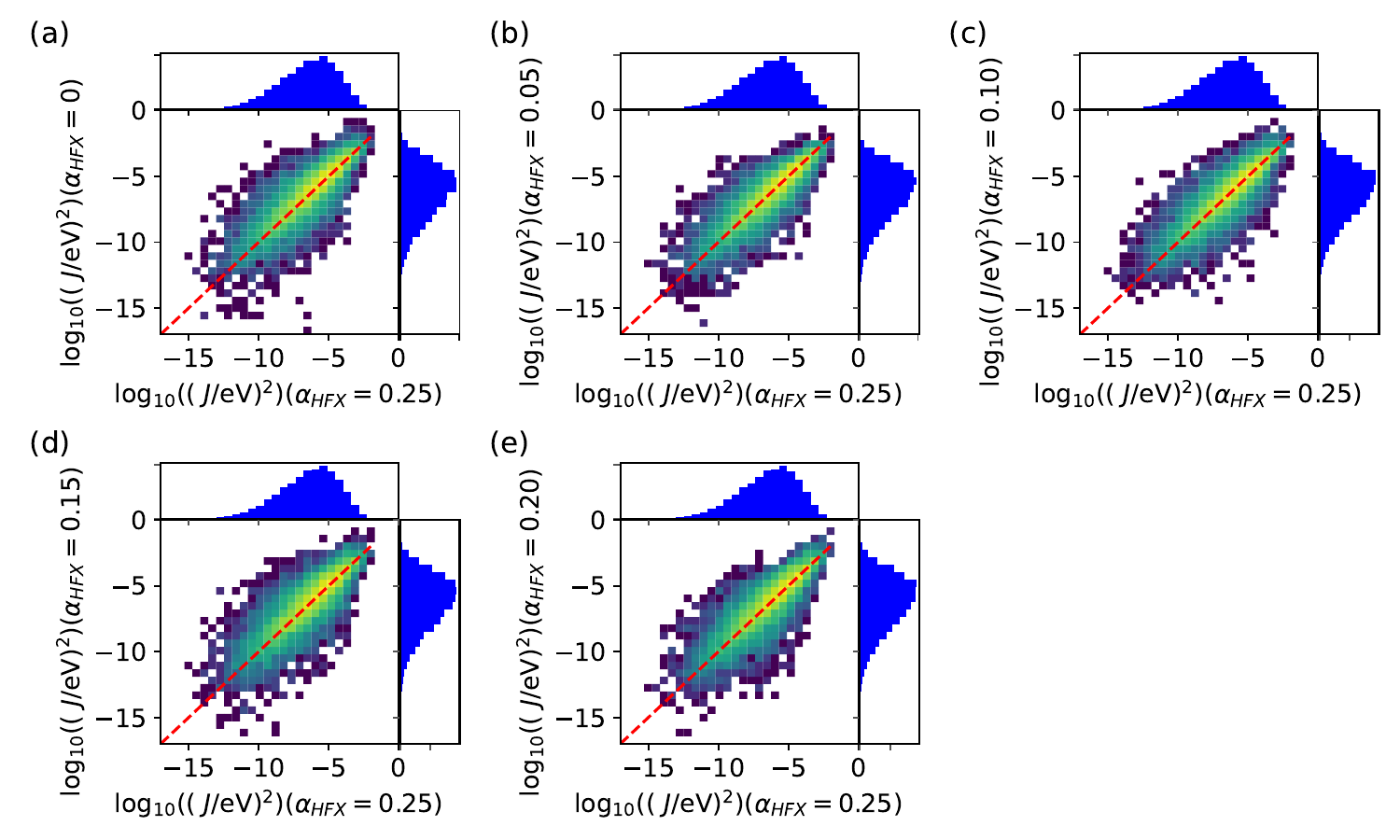}
  \caption{Scatter plot of electronic coupling elements of the MADN system for different values of \ahfx vs the PBE0 ($\ahfx=0.25$) reference.}
  \label{fig:J_MADN}
\end{figure*}

Distributions and correlations compared with the PBE0 reference for the electronic coupling elements calculated with different values for \ahfx are shown in Figure~\ref{fig:J_MADN}. The individual distributions appear very similar, with a peak of $\log_{10}[(J_{ij}/\unit[]{eV})^2]$ between \unit[-5]{} to \unit[-6]{}, and a long tail of the distribution towards more negative values. The comparison with the PBE0 reference shows that while there is a clear correlation between the results for different values of \ahfx, the spread in the order of magntitudes of $J_{ij}^2$ can be very large especially for the lower coupling regions. Overall, the squared electronic coupling elements are found in a very wide range from $\unit[10^{-2}]{}$ to $\unit[10^{-15}]{(eV)^2}$ due to its exponential distance dependence and sensitivity to mutual orientation of the two involved molecules. Whereas the site energy distributions discussed in the preceding section are well-defined in the sense that each energy is unambiguously associated to a physical entity -- a molecule in the morphology --, the coupling elements are evaluated for a neighborlist based on a chosen cutoff as explained in Section~\ref{sec:model}. Clearly, if this cutoff is chosen to be large, a lot of hopping pairs with very small coupling elements will be considered that may not be relevant at all (or even unphysical) for charge transport. Therefore, the increasing deviations for the most negative values in Figure~\ref{fig:J_MADN} may not be relevent either. 

\begin{figure}[tbp]
  \centering
  \includegraphics[width=\linewidth]{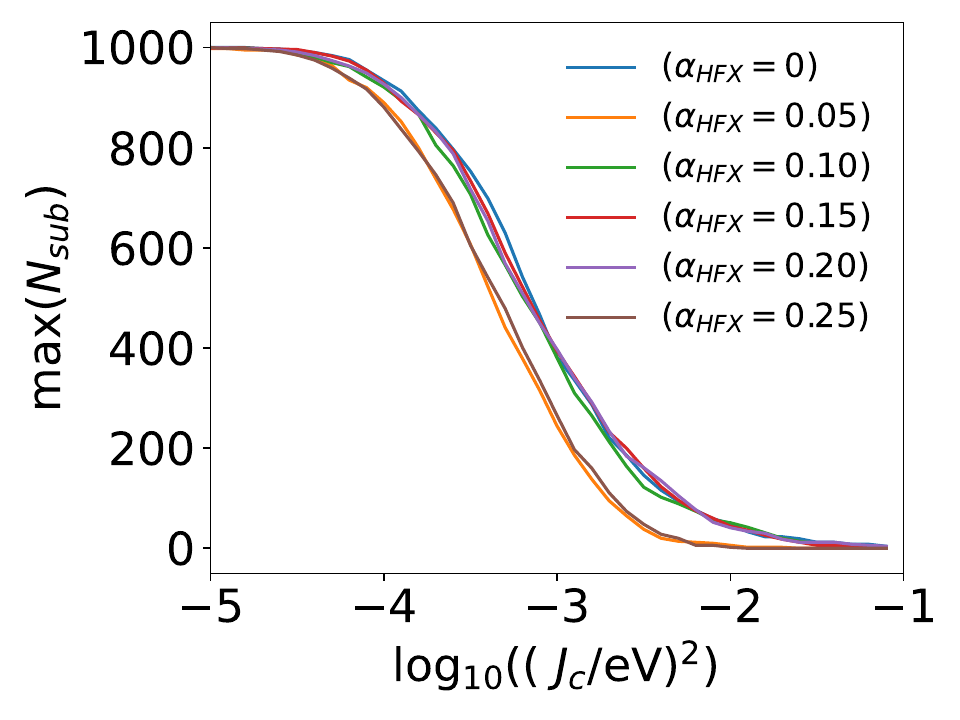}
  \caption{The maximum size of the subgraphs obtained in the percolation algorithm as a function of critical value $J_c$.}
  \label{fig:J_percolate}
\end{figure}

To determine the range of $\log_{10}[(J_{ij}/\unit[]{eV})^2]$ that are significant for the charge dynamics, a percolation analysis is performed to find a critical threshold value for the squared electronic coupling below which the largest connected subgraph is identical to the full graph. This is achieved by removing for a given value for $J_\text{c}$ from the full graph the edges with $J_{ij}^2 < J^2_\text{c}$ followed by the determination of the number of vertices in the largest connected subgraph $\max(N_\text{sub})$. Figure~\ref{fig:J_percolate} shows the resulting dependence of $\max({N_\text{sub}})$ on the critical value $J_c$. It is apparent that for $\log_{10} J_c^2 = -5$, all vertices are in the largest connected subgraph. The overall connectivity of the charge transport network is therefore mostly affected by the coupling elements larger than this threshold, and the charge transport properties are expected to be more sensitive to deviations for the associated edges. We also note that overall, the results in Fig.~\ref{fig:J_percolate} again seem to be very similar for all \ahfx studied in the work.

\subsection{Time-of-flight calculations}

\begin{table}[tbp]
\caption{Simulated ToF (in \unit[]{s}) for different values of \ahfx in the full graph, with an additional cutoff on $J^2$, and for the case with no energetic disorder, respectively.}
\begin{ruledtabular}
  \begin{tabular}{c c c c }
        \ahfx & full & $J_{ij} \mathbb{1}(\log_{10} J^2_{ij} > -5)$ & $\Delta E_{ij} =0$ \\
    \hline
        0.00 &  $6.4 \cdot 10^{-9}$ & $7.5 \cdot 10^{-9}$ & $1.9 \cdot 10^{-10}$ \\
        0.05 & $ 7.9 \cdot 10^{-9}$ & $8.1 \cdot 10^{-9}$ & $4.1 \cdot 10^{-10}$ \\
        0.10 & $ 1.6 \cdot 10^{-8}$ & $1.6 \cdot 10^{-8}$ & $4.0 \cdot 10^{-10} $ \\
        0.15 & $ 3.0 \cdot 10^{-8}$ & $3.1 \cdot 10^{-8}$ & $4.0 \cdot 10^{-10} $ \\
        0.20 & $ 2.1 \cdot 10^{-8}$ & $2.5 \cdot 10^{-8}$ & $4.5 \cdot 10^{-10}$ \\
        0.25 & $ 9.5 \cdot 10^{-8}$ & $9.7 \cdot 10^{-8}$ & $7.2 \cdot 10^{-10}$ \\
    \end{tabular}
\end{ruledtabular}
\label{tab:ToF_MADN_HFX}
\end{table}

From the analysis of reorganization energies, site energies, and electronic coupling elements for different values of \ahfx, it is not clear how the mostly on distribution-level observed variations impact the overall charge transport properties. To scrutinize the dependence of such a material property, we now calculate the time-of-flight $\tau$ and report the respective values in Table~\ref{tab:ToF_MADN_HFX}. Here, we refer with ''full'' to time-of-flight obtained for the as-calculated charge transport network. One can see that $\tau$ varies by roughly one order of magnitude between $\ahfx=0$ ($\unit[6.3\cdot 10^{-9}]{s}$) and $\ahfx=0.25$ ($\unit[9.5\cdot 10^{-8}]{s}$), with an almost monotonous increase. When the squared coupling elements with values below $\unit[10^{-5}]{(eV)^2}$ are set to zero ($J_{ij} \mathbb{1}(\log_{10} J^2_{ij} > -5)$), one obtains only minimally, but consistently, larger $\tau$, corroborating the notion that the very small coupling elements are of little relevance for charge transport. Finally, we also consider the case in which the energetic disorder is ignored ($\Delta E_{ij}=0$). Here, one also can see (next to the generally shorter time-of-flight) a consistent increase in $\tau$, however only by a factor of about \unit[3.8]{}, a combined effect of the increased reorganization energy and variations in coupling elements.

\section{Uncertainty Quantification and Sensitivity Analysis}
\label{sec:UQ}
The previous section shows that different HFX affect the calculated ToF. In this section, we use the Monte Carlo method to estimate the range of the ToF given a confidence level, followed by a sensitivity analysis to determine which parameter contributes more to the variance of ToF. 

\subsection{Time-of-flight of diffusion}
As discussed the ToF is calculated from the graph with weighted edges defined by the electronic structure data $(\lambda, E_1,\ldots,E_N, J_1^2,\ldots,J_{N_\text{p}}^2)$, where $N$ is the number of molecules (vertices) and $N_\text{p}$ the number of molecule pairs (edges) in the neighbor list. We can therefore consider that $\tau = \tau(\mathbf{x})$ with $\mathbf{x}=(\lambda, E_1,\ldots,E_N, \log_{10}(J_1^2),\ldots,\log_{10}(J_{N_\text{p}}^2))$ as dependent on $M = 1+N+N_\text{p}$ parameters with uncertainty, in this case stemming from the possible choices of \ahfx in the exchange-correlation functional of the multiscale model. Note that due to the data range for the electronic coupling element as discussed in the previous section, we consider $\log_{10}(J^2)$ instead of $J^2$. For MADN, we have $N=1000$ and $N_\text{p}=8764$ so that $M=9765$.

Because the precise distribution of the uncertainties of the $M$ parameters is not known, we consider the maximum amount of uncertainties from a maximum likelihood estimation based on the data obtained from the explicit results for the different values of \ahfx and the assumption of normal distributions. In other words, each of the parameters $x_i$ for $i=1,\ldots,M$ is assigned a normal distribution with $\mathcal{N}(\mathbb{E}(x_i),\mathbb{V}(x_i))$, with $\mathbb{E}(x_i)$ ($\mathbb{V}(x_i)$) the mean (variance) of the respective data. Then, Monte-Carlo sampling is used to obtain $N_\text{MC}=50000$ different realizations of $\mathbf{x}$, the respective time-of-flight is calculated from these samples, and the resulting distribution $P(\log_{10}(\tau))$ is statistically analyzed. Specifically, we consider four different settings: in three settings, only one of the parameters blocks $(\lambda,\left\{E\right\},\left\{\log_{10}(J^2)\right\})$ is sampled while the values of the other blocks are set to their respective mean values, and we denote the resulting distributions as $P_\lambda(\log_{10}(\tau))$, $P_E(\log_{10}(\tau))$, and $P_J(\log_{10}(\tau))$, respectively. In the fourth setting, all parameters are sampled at the same time, yielding $P_\mathbf{x}(\log_{10}(\tau))$. As these distributions do not necessarily follow any specific analytic form, we estimate a $99\%$ confidence interval around the median using the equal-tailed percentile method. 

\begin{figure}[tbp]
  \centering
  \includegraphics[width=0.45\textwidth]{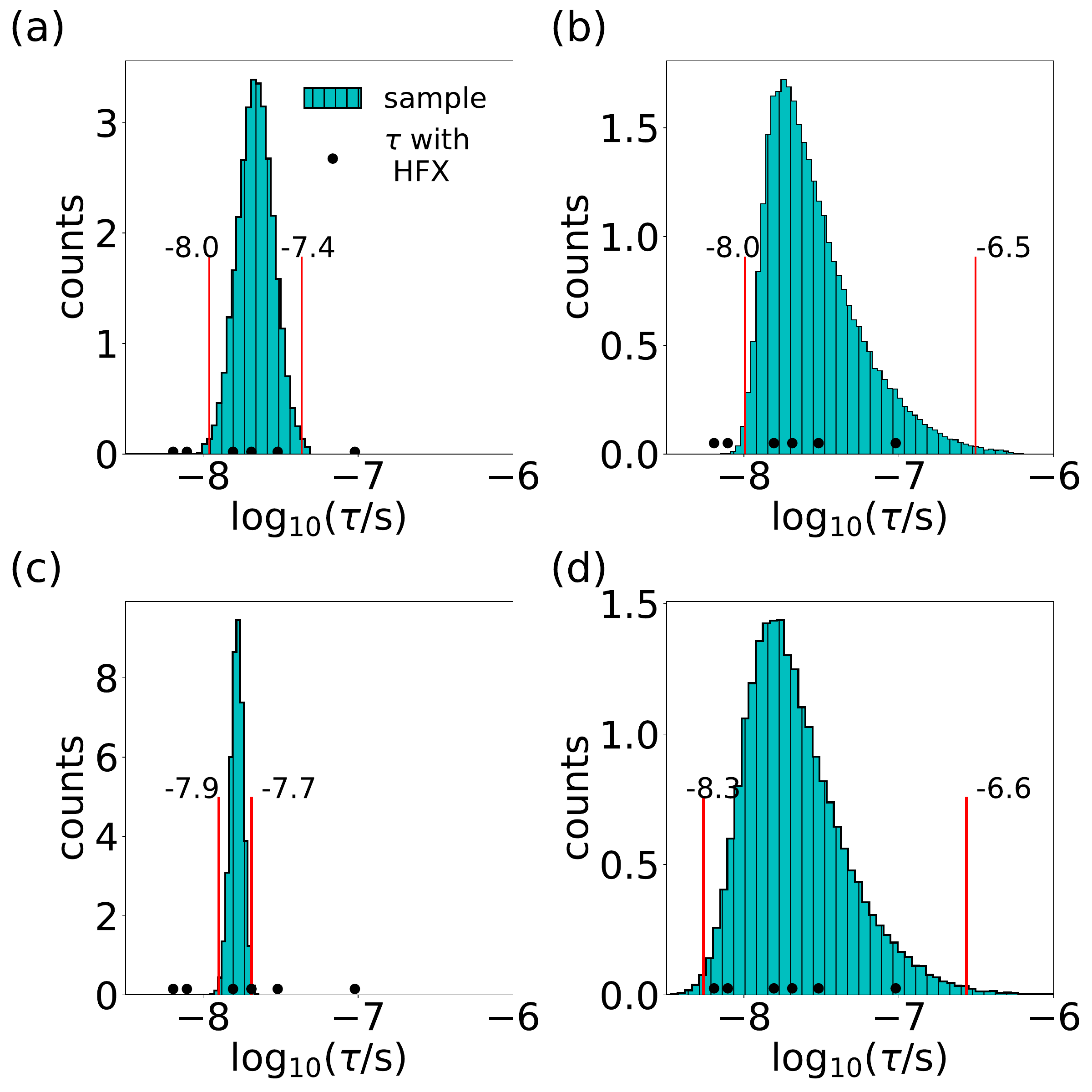}
  \caption{Distribution of ToFs in the MADN system from  Monte Carlo sampling with a sample size $N_\text{MC}=50000$. The red vertical lines indicate the lower bound and upper bound of the 99\% confidence interval around the median. The black circles indicate the ToF obtained using explicit \ahfx values. Uncertainties considered are (a) in the eorganziation energy, (b) in the site energy, (c) in the coupling elements, and (d) in all simultaneously.}
  \label{fig:mle_MADN_withE}
\end{figure}

The four respective distributions of $\tau$ in the MADN system are shown in Fig.~\ref{fig:mle_MADN_withE} together with the indicated confidence intervals obtained from Monte-Carlo sampling. Black dots at the $x$-axes are the data from the explicit calculations for the six \ahfx values discussed before. Considering first the three distributions with single property sampling in panels (a)-(c), we find the confidence intervals for $\log_{10}(\tau)$ of $[-8.0,-7.4]$ for uncertain $\lambda$, $[-8.0,-6.5]$ for uncertain $E$, and $[-7.9,-7.7]$ for uncertain $J$, respectively. Interestingly, all three share very similar lower limits of the confidence interval, while the upper limits and the shape of the distributions differ significantly. While $P_\lambda(\log_{10}(\tau))$ and $P_J(\log_{10}(\tau))$ appear relatively symmetric, indictaing possible a log-normal distribution of $\tau$, $P_E(\log_{10}(\tau))$ is distinctly asymmetric which might point to a normal distribution of $\tau$ for uncertain energies. From the three distributions and the confidence intervals one can see that the effect of the uncertainty in the coupling elements on the time-of-flight is small (factor of 2 on $\tau$). The effect of uncertainty in the reorganization energy is only slightly higher (factor of four). Uncertainty in the site energies, in contrast, has a much bigger impact on the estimates of $\tau$. From the distribution of $P_\mathbf{x}(\log_{10}(\tau))$ in Fig.~\ref{fig:mle_MADN_withE}(d), one can see than when all parameters are sampled, the result is dominated by the effects from the uncertainty in the site energies, with an overall similar shape and a confidence interval of $[-8.3,-6.6]$, corresponding to a range from about \unit[5]{ns} to about \unit[250]{ns} in $\tau$. 

We note that the above analysis from sampling via independent distributions of the uncertain parameters assumes that the parameters themselves are independent or uncorrelated. Correlations in organic semiconductors are typically found among the site energies when the molecules are polar~\cite{Baumeier2011,10.1063/5.0049513}. MADN has a very low dipole moment, and we find practically no spatial correlations between the site energies, making the sampling approach suitable for this material system.

Besides the inspection on the distributions for various uncertain parameter (sets) as above, one can also quantify the sensitivity of the time-of-fight to the respective parameters by evaluating their contributions to the overall observed variance in $\tau$. One way of decomposing the variance of the model output into fractions attributed to input parameters is the variance-based sensitivity analysis using Sobol indices~\cite{saltelli_variance_2010}. 

To measure the parameter $x_i$'s contribution to $\tau=\tau(\mathbf{x})$ including all variance caused by its interaction with other parameters $\{x_k, k \neq i \}$, the total Sobol index $S_{T,i}$ is calculated as
\begin{equation}
    S_{T,i} = \frac{ \mathbb{E}_{\mathbf{x}_{\sim i}}[ \mathbb{V}_{x_i}(\tau|\mathbf{x}_{\sim i}) ] }{ \mathbb{V}(\tau) } .
    \label{eq:STi}
\end{equation}
Here, $\mathbf{x}_{\sim i}$ denotes the vector of all entries of $\mathbf{x}$ but $x_i$. The variance of $\tau$ given a set of $\mathbf{x}_{\sim i}$ taken over $x_i$ is $\mathbb{V}_{x_i}(\tau|\mathbf{x}_{\sim i})$ and $ \mathbb{E}_{\mathbf{x}_{\sim i}}[\cdot]$ denotes the mean of argument $(\cdot)$ taken over all factors but $x_i$. We then consider based on the structure of the vector $\mathbf{x}$ the Sobol index for the reorganization energy as $S_{T,\lambda} = S_{T,1}$, for the site energies as $S_{T,E} = \sum\limits_{i=2}^{N+1} S_{T,i}$, and for the coupling elements as $S_{T,E} = \sum\limits_{i=N+2}^{N+1+N_\text{p}} S_{T,i}$. Using the quasi Monte Carlo method\cite{sobol_global_2001} with a sample size $N_\text{QMC}=1000$ to calculate $S_{T,i}$, we find that $S_{T,\lambda}=0.097$, $S_{T,E} =0.950$ and $S_{T,J}=0.028$. The Sobol indices corroborate the observation from the confidence intervals that the uncertainty in the site energies gives the dominant contribution to the variance of the simulated time-of-flight. Contributions from the reorganization energy and the coupling elements are both small in relation, with the one of $\lambda$ slightly larger than the one of $J$. This is also in line with the general observations on the distributions in Fig.~\ref{fig:mle_MADN_withE}.

\subsection{Drift-diffusion Charge mobility}
When an external electric field, $\mathbf{F}$, is applied to the system, the transport is not just diffusive but a drift-diffusion process. The material property of interest is then the drift mobility
\begin{equation}
    \mu = \frac{\mathbf{v} \mathbf{F} }{ |\mathbf{F}|^2},
    \label{eq:mu}
\end{equation}
where $\mathbf{v} = \mathbf{s}/{\tau}$ is the effective velocity of the charge carrier and $\mathbf{s}$ is the vector connecting the initial and final positions of the charge carrier. Here, we perform calculations for six different field directions (positive and negative Cartesian directions) with the source and sink conditions set accordingly. The electric field is taken account in the Marcus rate in Eq.~\ref{equ:Marcus} via an additional term in $\Delta E_{ij}$ given by $e\mathbf{F} \cdot \mathbf{r}_{ij}$, where $\mathbf{r}_{ij}$ connects the center-of-masses of molecules $i$ and $j$. Finally, we define for a certain field strength $F=\vert\mathbf{F}\vert$ the ToF drift mobility $\mu(F)$ is as the average over the six directions. 

\begin{figure}[tbp]
    \centering
    \includegraphics[width=\linewidth]{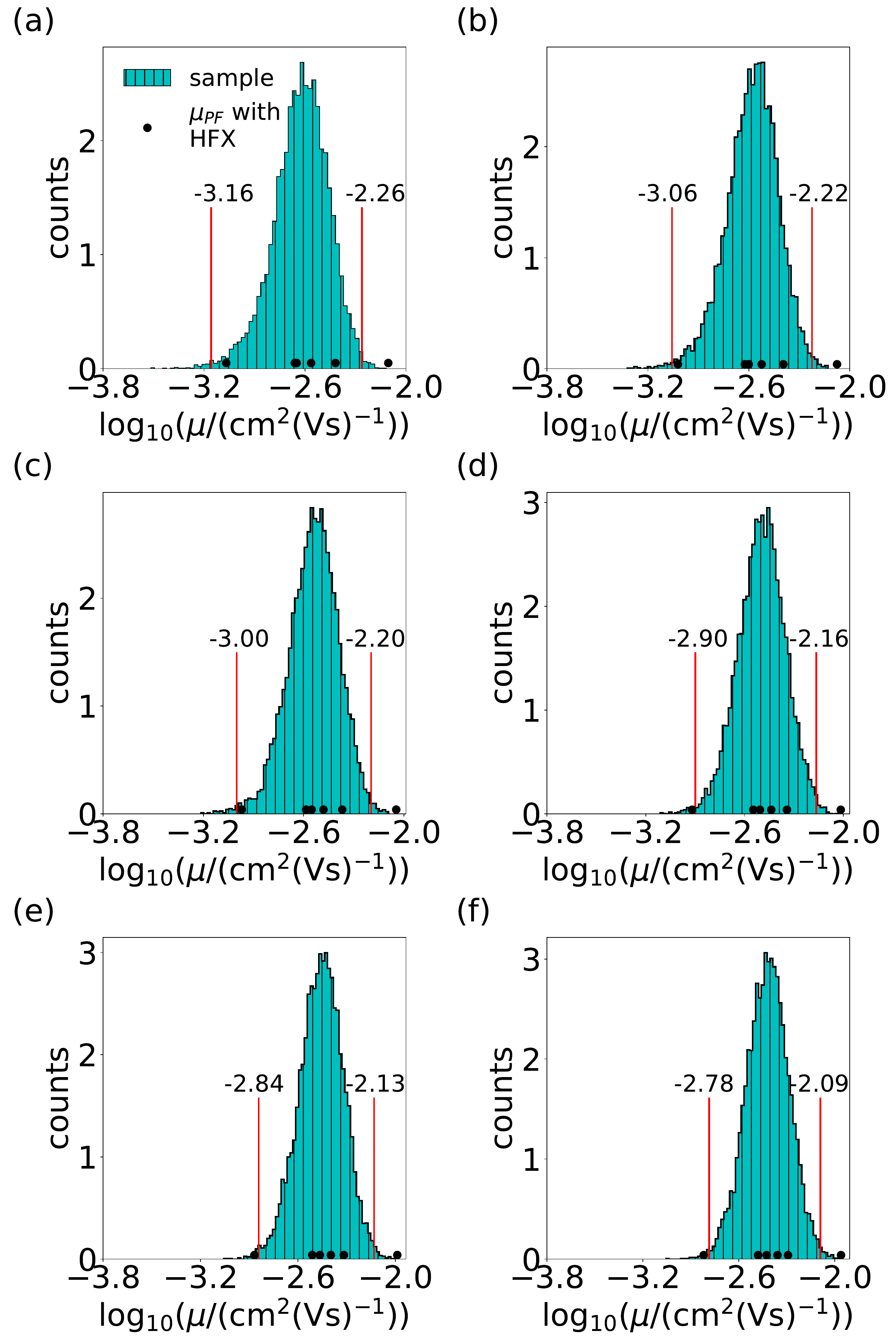}
    \caption{Distribution of the mobility $\mu$ for different values of the externally applied electric field $F$ as obtained from uncertainty quantification using 
    Monte Carlo sampling with a sample size $N_\text{MC}=10000$. The red vertical lines indicate the lower bound and upper bound of the 99\% confidence interval around the median. (a) $|\vec{F}|=4 \cdot 10^7 \unit{V/m}$, (b) $|\vec{F}|=5 \cdot 10^7 \unit{V/m}$, (c) $|\vec{F}|=6 \cdot 10^7 \unit{V/m}$, (d) $|\vec{F}|=7 \cdot 10^7 \unit{V/m}$, (e) $|\vec{F}|=8 \cdot 10^7 \unit{V/m}$, (f) $|\vec{F}|=9 \cdot 10^7 \unit{V/m}$.}
    \label{fig:fig_mle_withE_mu2_ave}
\end{figure}

We perform such mobility calculations for the six explicit \ahfx models as well as the Monte-Carlo sampled uncertainty for six different values of $F$. The results are shown in Fig.~\ref{fig:fig_mle_withE_mu2_ave}. While the distributions shift depending on the strength of the applied field, the overall shape and the width of the confidence intervals is very similar. The width of the confidence intervals in $\log_{10}(\mu)$ decreases from \unit[0.9]{} at low fields ($\unit[4 \cdot 10^7]{V/m}$) to \unit[0.7]{} at a field of $\unit[9 \cdot 10^7]{V/m}$, corresponding broadly to a width of close to one order of magnitude in $\mu$. A stronger drift component due to a larger external electric field seems to make the mobility slightly less sensitive to uncertainties. 

\begin{figure}[tbp]
    \centering
    \includegraphics[width=0.45\textwidth]{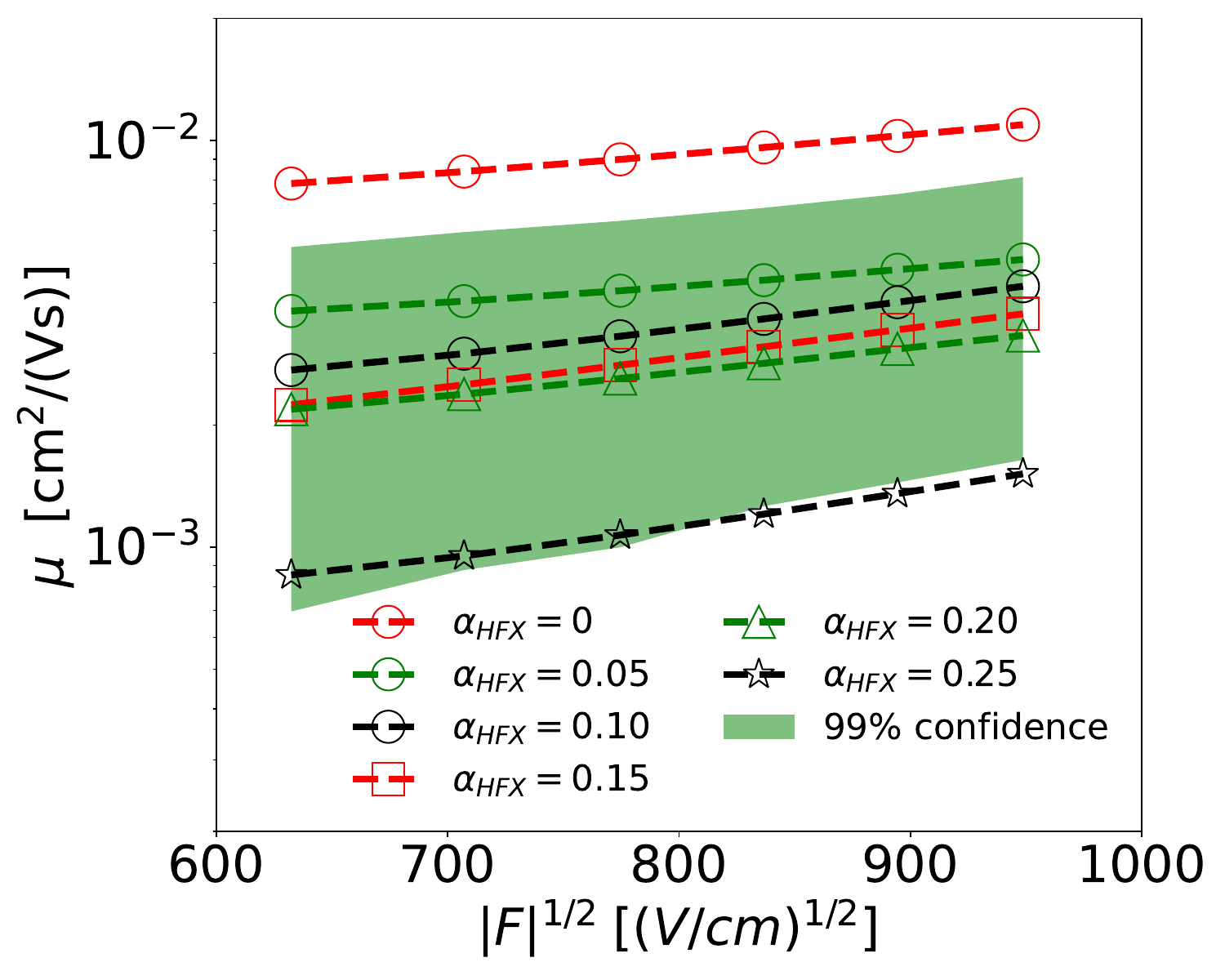}
    \caption{Electric-field dependence of the mobility $\mu$ in the multiscale modeled MADN system. The dash lines are Poole--Frenkel plots obtained with specific \ahfx values, and the green shaded area indicates the 99\% confidence interval estimated from the MC sampling with a sample size of $N_\text{MC}=10000$.}
    \label{fig:PF_plot_ave}
\end{figure}

\begin{table}[tbp]
  \caption{\label{tab:PF_parameter}%
  Poole-Frenkel parameters $\mu_0$ (in $\unit[]{cm^2/(Vs)}$) and $\beta$ (in $\unit[]{\sqrt{cm/V}}$) of the multiscale modeled MADN, calculated by the six different \ahfx values. 
  }
  \begin{ruledtabular}
    \begin{tabular}{c c c}
    $\ahfx$ & $\mu_0$  & $\beta$ \\
      \hline
    0 & $4.0 \cdot 10^{-3}$ & $1.0 \cdot 10^{-3}$ \\
    0.05 & $2.1 \cdot 10^{-3}$ & $9.3 \cdot 10^{-4}$ \\
    0.10 & $1.0 \cdot 10^{-3}$ & $1.5 \cdot 10^{-3}$ \\
    0.15 & $8.0 \cdot 10^{-4}$ & $1.6 \cdot 10^{-3}$ \\
    0.20 & $9.3 \cdot 10^{-4}$ & $1.3 \cdot 10^{-3}$ \\
    0.25 & $2.6 \cdot 10^{-4}$ & $1.8 \cdot 10^{-3}$ \\
      \end{tabular}
  \end{ruledtabular}
  \end{table}

Figure~\ref{fig:PF_plot_ave} shows the plot of the charge mobility one logarithmic $y$-axis as a function of $\sqrt{F}$ to reveal the electric field dependence of the type $\mu(F)=\mu_0 \exp (\beta \sqrt{F})$ as predicted by Poole and Frenkel~\cite{frenkel_prebreakdown_1938}. We show the explicit results for the six chosen \ahfx values together with the confidence interval estimate from the uncertainty quantification via Monte-Carlo sampling. In Table~\ref{tab:PF_parameter} we also summarize the  parameters $\mu_0$ and $\beta$ extracted from fits of the explicit data to the Poole--Frenkel expression. The values for $\mu_0$ vary from $\unit[2.6\cdot 10^{-4}]{cm^2/(Vs)}$ to $\unit[4.0\cdot 10^{-3}]{cm^2/(Vs)}$, roughly one order or magnitude depending on the chosen \ahfx, in line with the uncertainty from the Monte-Carlo sampling. The Poole--Frenkel slope $\beta$ in contrast is found to be much less sensitive to uncertainties, with recorded values between $\unit[9.3\cdot 10^{-4}]{\sqrt{cm/V}}$ and $\unit[1.8\cdot 10^{-3}]{\sqrt{cm/V}}$.

\section{Discussion \& Conclusion}
One key observation of this study of Uncertainty Quantification in multiscale models for charge transport in organic semiconductors due to the choice of exchange-correlation functionals in density functional theory by systematically varying the Hartree--Fock exchange fraction in hybrid functionals is the dominance of site energy uncertainties in influencing the variability of the time-of-flight and mobility predictions. In contrast, the uncertainties in reorganization energies and electronic coupling elements are shown to have a relatively minor effect. This suggests that further efforts to enhance the accuracy of site energy calculations -- potentially through improved parameterization or advanced quantum mechanical methods -- could significantly improve the reliability of multiscale modeling predictions.   Interestingly, the results also demonstrate that the uncertainties in electronic coupling elements, particularly those with very small magnitudes, are less relevant for overall charge transport. This aligns with the percolation analysis, which indicates that coupling elements below a critical threshold contribute minimally to the connectivity of the charge transport network. Thus, simplifying approximations in handling these low-value couplings could be justified in specific scenarios to reduce computational complexity without compromising accuracy.  

Another notable finding is the robustness of field-dependent mobility predictions. Despite uncertainties in the underlying molecular parameters, the electric field dependence of mobility exhibits relatively consistent Poole--Frenkel behavior across the tested range of \ahfx values. This robustness suggests that multiscale models retain predictive value for field-dependent trends, even when parameter uncertainties are present. However, the results also highlight the challenge of achieving quantitative accuracy in charge mobility predictions, given that the uncertainty range spans an order of magnitude. This raises important considerations for interpreting simulation results in the context of experimental data. For instance, small differences in predicted mobility between two materials or designs may not be statistically significant given the intrinsic uncertainties.  

Despite the progress made, there are limitations to this study that warrant further investigation. First, the analysis is restricted to variations in \ahfx as a proxy for different exchange-correlation functionals. While this approach captures a significant portion of the uncertainty, other factors such as basis set selection, dispersion corrections, and model approximations may also contribute to variability in the results. Expanding the uncertainty quantification framework to include these factors would provide a more comprehensive understanding of the robustness of multiscale models.  

Furthermore, the study focuses on a single material system, MADN, which exhibits moderate disorder. Extending the analysis to other OSCs with varying degrees of disorder, charge-carrier types, or different molecular morphologies would enhance the generality of the conclusions.

\appendix*
\section{Electrostatic and polarized energy contributions}
\begin{figure*}
  \centering
  \includegraphics[width=0.80\textwidth]{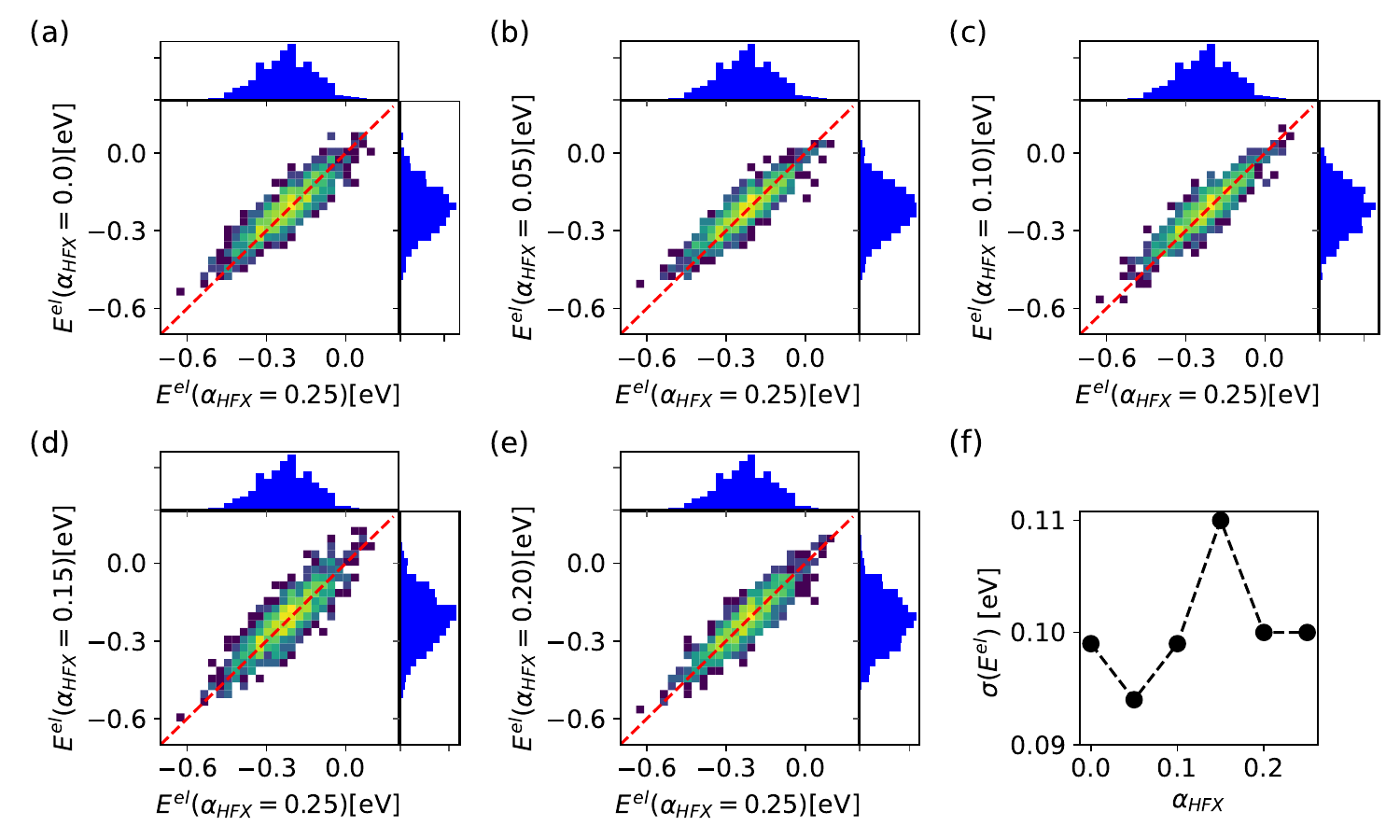}
  \caption{Panels (a)-(e): Scatter plot of the electrostatic contribution to site energies of the MADN system for different values of \ahfx vs the PBE0 ($\ahfx=0.25$) reference. (f) Site energy disorder $\sigma$ in \unit[]{eV} for all values of \ahfx.}
  \label{fig:Estat_qmmm_MADN}
\end{figure*}

\begin{figure*}
  \centering
  \includegraphics[width=0.80\textwidth]{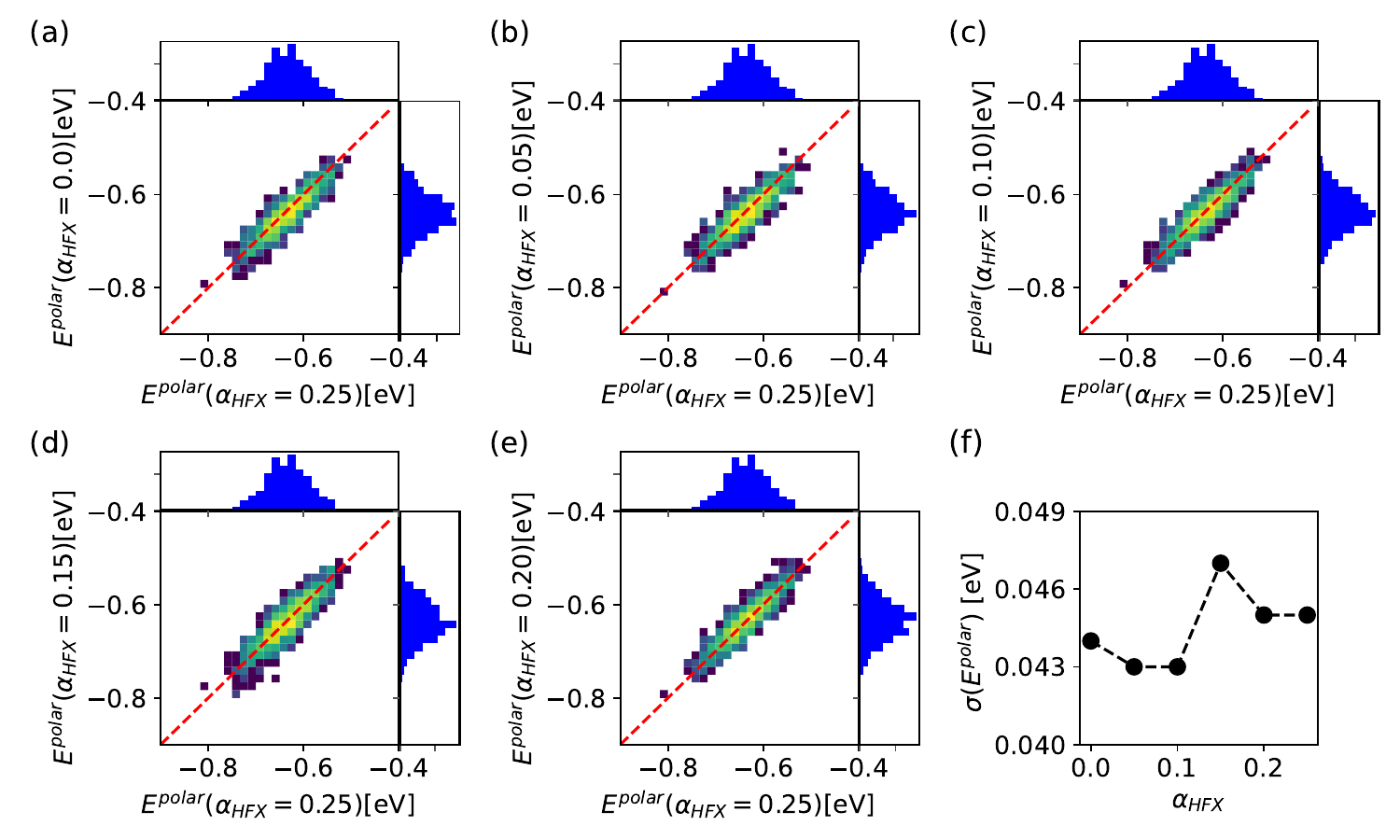}
  \caption{Panels (a)-(e): Scatter plot of polarization contribution to site energies of the MADN system for different values of \ahfx vs the PBE0 ($\ahfx=0.25$) reference. (f) Site energy disorder $\sigma$ in \unit[]{eV} for all values of \ahfx.}
  \label{fig:Edip_qmmm_MADN}
\end{figure*}

\end{document}